\documentclass{rmf-d}
\usepackage{nopageno,rmfbib,multicol,times,epsf,amsmath,amssymb,cite,float}
\usepackage[latin1]{inputenc}
\usepackage{caption2}
\usepackage{graphicx}
\usepackage{gensymb}

\def\rmfcornisa{Research OR Education of Physics \hfill\rmf\ {\bf ??} (*?*) ???--???
\hfill MES? A\~NO?}

\def\rmfcintilla{{\it Rev.\ Mex.\ Fis.\/} {\bf ??} (*?*) (????) ???--???}
\clearpage \rmfcaptionstyle 
\begin{document}
\title{On the mass of the glueballonium    
\vspace{-6pt}}
\author{ Enrico Trotti$^{1}$ and Francesco Giacosa$^{1,2}$  }
\address{$^{1}$ Institute of Physics, Jan Kochanowski University, ul. Uniwersytecka 7, 25-406, Kielce, Poland \\
$^{2}$ Institute for Theoretical Physics, J. W. Goethe University, Max-von-Laue-Str. 1, 60438 Frankfurt, Germany}
\author{ }
\address{ }
\author{ }
\address{ }
\author{ }
\address{ }
\author{ }
\address{ }
\author{ }
\address{ }
\maketitle
\recibido{day month year}{day month year
\vspace{-12pt}}
\begin{abstract}
\vspace{1em}According to lattice simulations and other theoretical approaches, the scalar glueball is the lightest state in the Yang-Mills sector of  QCD. Since within this sector the scalar glueball is stable, the scattering between two glueballs is a well-defined process. Moreover, a glueball-glueball bound state, called glueballonium, might exist if the attraction turns out to be large enough.  In this work, we concentrate on the formation of the glueballonium in the context of the dilaton potential. In particular, we investigate the parameter values for which such a glueballonium emerges.
\vspace{1em}
\end{abstract}
\keys{Hadrons, glueballs, bound state  \vspace{-4pt}}
\pacs{   \bf{\textit{14.40.Rt, 11.80.Et, 12.39.Mk }}    \vspace{-4pt}}
\begin{multicols}{2}

\section{Introduction}
In the Yang-Mills (YM) sector of Quantum Chrmodynamics (QCD), glueballs, i.e. bound states of gluons, have been predicted by a variety of approaches, which range from bag models \cite{Chodos:1974je,Jaffe:1975fd,Jaffe:1985qp} to lattice QCD \cite{Morningstar:1999rf,Chen:2005mg,Caselle:2001im}. 
In all these approaches, the lightest glueball has scalar quantum numbers $J^{PC}=0^{++}$ and is therefore stable in the YM theory. Hence, the scattering between two scalar glueballs is a well-defined process that can be investigated in models and on the lattice. 
Here, following the discussion of Ref. \cite{Giacosa:2021brl}, we study the scattering of two glueballs by using the well-known dilaton potential \cite{Migdal:1982jp,Salomone:1980sp}. Moreover, if the attraction is strong enough, a glueball-glueball bound state, referred to as the glueballonium, may form.  
In these proceedings, we study under which conditions the dilaton potential allows for the emergence of the glueballonium. To this end, one needs to introduce a unitarization of the scattering amplitude: a rather simple and often used unitarization within the so-called on-shell approximation is applied. 

Quite interestingly, such a glueballonium could be searched for on lattice YM and, eventually, in experiments.  For the latter, it is important to remind that in full QCD the scalar glueball (as any other glueball) is not stable.  For instance, a good candidate for being predominantly the scalar glueball is e.g. $f_0(1710)$ \cite{Janowski:2014ppa,Close:2001ga,Brunner:2015oqa,Rodas:2021tyb,Giacosa:2005qr}, which decays predominantly into kaons and pions. As a consequence, the glueballonium, even if existent and stable in YM, would also not be stable in full QCD and would decay into two and four conventional mesons (such as pions, kaons, etc.). Besides the necessary existence in the YM sector, its eventual discovery is therefore  possible only if it turns to be narrow enough.

\section{Tree-level scattering}
A quite famous low-energy theory of the YM sector of QCD is given by the dilaton Lagrangian, which contains a single scalar dilaton/glueball field $G$  \cite{Migdal:1982jp,Salomone:1980sp,Ellis:1984jv}:
\begin{equation}%
\mathcal{L}%
_\text{dil}=\frac{1}{2}(\partial_{\mu}G)^{2}-V(G),
\label{lagra}
\end{equation}
with%
\begin{equation}
V(G)=\frac{1}{4}\frac{m_{G}^{2}}{\Lambda_{G}^{2}}\left(  G^{4}\ln\left\vert
\frac{G}{\Lambda_{G}}\right\vert -\frac{G^{4}}{4}\right)  \text{.}%
\label{potential}%
\end{equation}
The dilaton potential contains one dimensional parameter, the scale $\Lambda_{G}$, which embodies the trace anomaly at the composite confined level, see e.g \cite{Ellis:1984jv}.  Moreover, the dilaton potential contains the dimensionless quantity $m_G/\Lambda_{G}$. The quantity $m_G$ corresponds to the glueball mass (second derivative at the minimum $G=\Lambda_{G}$); its numerical value reads $m_G \approx 1.7$ GeV obtained in lattice QCD \cite{Morningstar:1999rf,Chen:2005mg,Caselle:2001im}. The scale $\Lambda_{G}$ can be obtained by a comparison with the gluon condensate and takes the (approximate) value $\Lambda_{G}\approx0.4$ GeV \cite{DiGiacomo:2000irz}, but this determination is quite uncertain, see e.g. Ref. \cite{Janowski:2014ppa}. Indeed, one of the goal of the glueball-glueball scattering would be an independent determination of this important quantity that affects the whole low-energy phenomenology \cite{Janowski:2014ppa,Carter:1995zi,Parganlija:2012fy,Parganlija:2010fz}.  

The total tree-level amplitude for the scattering $G(p_1)G(p_2) \rightarrow G(p_3)G(p_4)$ can be obtained from the 3- and 4-leg vertices obtained in the expansion of the potential around its minimum. As function of the Mandelstam variables $s = (p_{1} + p_{2})^{2}$, $t=(p_{1}-p_{3})^{2}$ and $u=(p_{2}-p_{3})^{2}$ the amplitude reads:
\begin{align}
A(s,t,u)=-11\frac{m_{G}^{2}}{\Lambda_{G}^{2}}-\left(  5\frac{m_{G}^{2}%
}{\Lambda_{G}}\right)  ^{2}\frac{1}{s-m_{G}^{2}} \nonumber \\
-\left(  5\frac{m_{G}^{2} 
}{\Lambda_{G}}\right)  ^{2}\frac{1}{t-m_{G}^{2}}   -\left(  5\frac{m_{G}^{2}%
}{\Lambda_{G}}\right)  ^{2}\frac{1}{u-m_{G}^{2}} \text{ .}\label{totampl}%
\end{align}
Since $s+t+u=4m_G^2$, only two of them are independent. Moreover, upon introducing the scattering angle $\theta$ and expanding in partial waves, we rewrite Eq. (\ref{totampl}) as:
\begin{equation}
    A(s,\cos \theta)=\sum_{\ell=0}^{\infty}(2\ell+1)A_{\ell}(s)P_{\ell}(\cos\theta) \text{ ,}
\end{equation}
where $P_{\ell}(\cos\theta)$ is the $l$-th Legendre polynomial. Conversely, the $l$-th amplitude reads:
\begin{equation}
    A_{l}(s)=\frac{1}{2}\int_{-1}^{1}d\cos\theta A(s,\cos\theta)P_{l}(\cos\theta) \text{ .}\label{l-ampl}
\end{equation}
Odd waves vanish because $A(s,\cos \theta)$ is symmetric in $\cos \theta$ for Bose symmetry. In these proceedings, we shall concentrate only on the S-wave  ($l=0$), which is the channel in which the glueballonium eventually forms (for the D-wave and G-wave, see Ref. \cite{Giacosa:2021brl}). The corresponding S-wave amplitude is given by:
\begin{equation}
A_{0}(s)=-11\frac{m_{G}^{2}}{\Lambda_{G}^{2}}
-25\frac{m_{G}^{4}}{\Lambda_{G}^{2}} \left( \frac{1}{s_{1}}
-2
\frac{\log\left(  \frac{s_{3}}{m_{G}^{2}}\right)  }{s_{4}}\right),
\label{eq:A0}
\end{equation}
where $s_{N}=s-N m_G^2$. 

The tree-level scattering length is calculated as:
\begin{equation}
     a_0 = \frac{ A_0(4m_G^2)}{32\pi m_G }{\equiv}
     \frac{1}{32\pi m_{G}}\frac{92m_{G}^{2}}{3\Lambda_{G}^{2}} \text{ ,}
     \label{length}
\end{equation}
which is inversely proportional to the energy scale $\Lambda_{G}^{2}$. A future determination of the scattering length determined on the lattice would be very useful:  since $m_G$ is known, one could use that determination in order to extract $\Lambda_{G}$. Yet, care is needed with our derivation: a unitarization is necessary since the tree-level result is not sufficient. In particular, it cannot access the eventual existence of a bound state. 

The unitarization process that we introduce below is constructed in such a way to leave unchanged the singularities of the tree-level amplitude at $s= m_G^2$ and $s=3m_G^2$. Namely, the pole in $m_G^2$ correspond to the single glueball state, so it is reasonable to require that it is left unchanged. Moreover, the $t-$ and $u-$channels projected onto the S-wave give rise to the left hand cut and, consequently, to a logarithmic singularity that we shall require to not be modified by the employed unitarization process.

\section{Unitarization and mass of the glueballonium}

\begin{figure}[H]
    \centering
    \includegraphics[width=\linewidth]{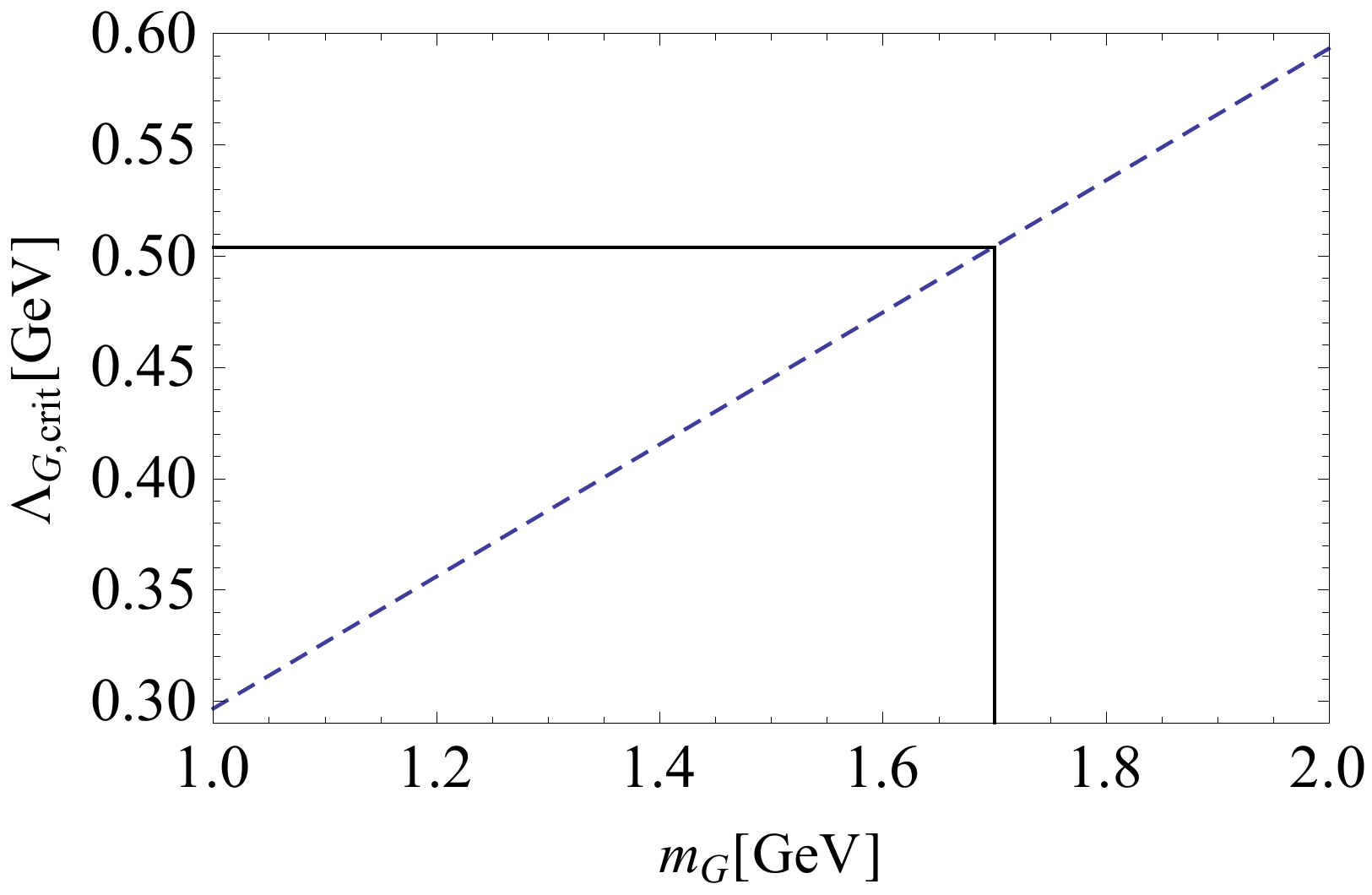}
    \caption{The critical value of the scale $\Lambda_{G,crit}$ as function of $m_G$; the value for $m_{G}=1.7$ is highlighted.}
    \label{figlin}
\end{figure}

\begin{figure}[H]
    \centering
    \includegraphics[width=\linewidth]{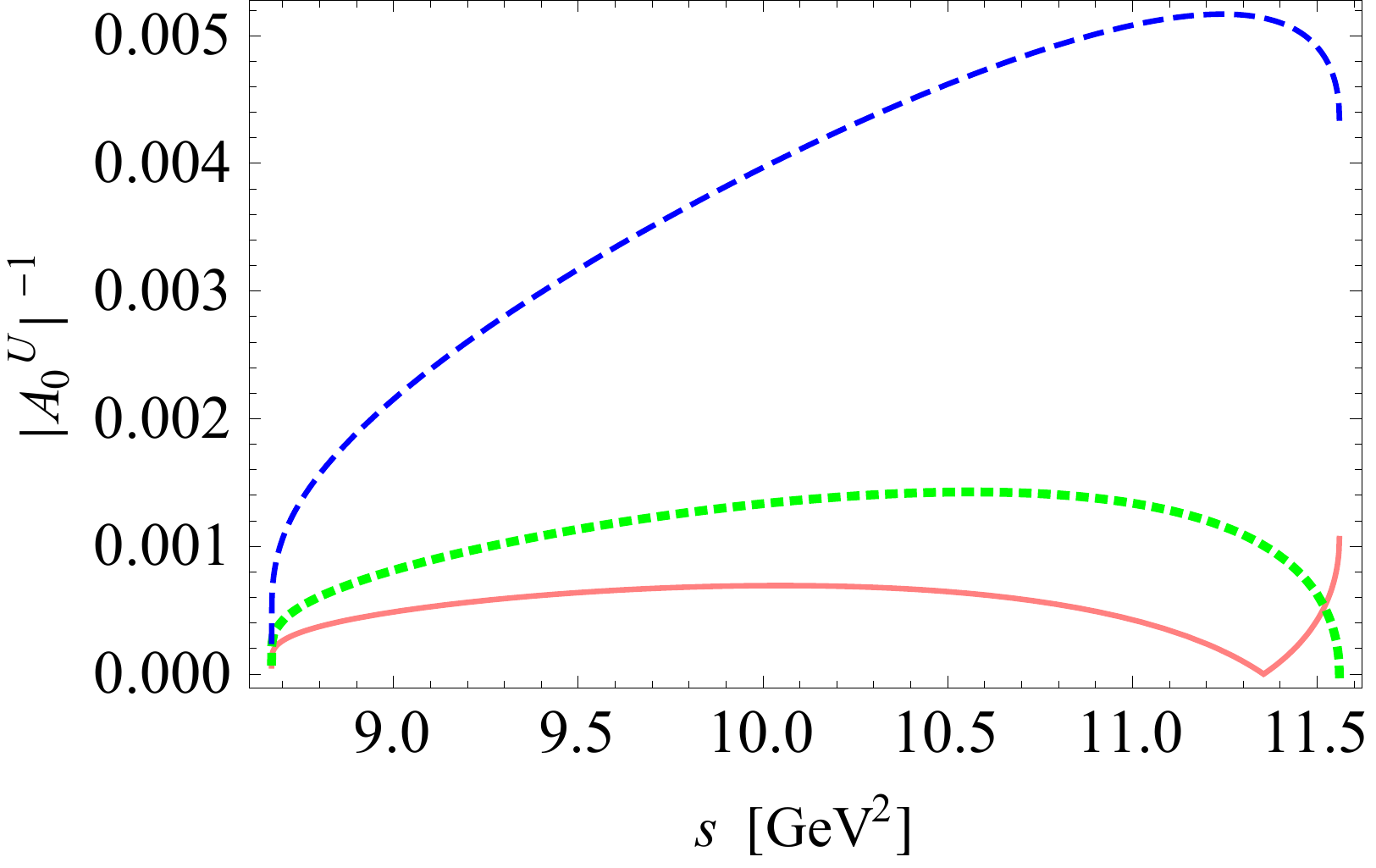}
    \caption{The inverse amplitude $|A^U_0|^{-1}$ as function of $s$, evaluated for three different values of $\Lambda$: $\Lambda_G=0.4$ GeV (pink, solid), $\Lambda_G = \Lambda_{G,\text{crit}} \approx 0.504$ GeV (green, dot), and $\Lambda_G=0.8$ GeV (blue, dash).}
    \label{figampl}
\end{figure}

Loop contributions have to be taken into account in order to produce poles representing bound states. The inclusion of these quantum fluctuations can be done through a proper unitarization of the tree-level partial wave amplitude. Among the various schemes proposed for chiral Lagrangians \cite{Truong:1991gv,GomezNicola:2001as,Selyugin:2007jf,Cudell:2008yb,Nebreda:2011di,Delgado:2015kxa,Oller:2020guq}, the one we choose here is the so-called on-shell approximation \cite{Dobado:1992ha,Oller:1997ng} leading to:
\begin{equation}
A_{0}^{\text{U}}(s)=\left[  A_{0}^{-1}(s)-\Sigma(s)\right]  ^{-1} \text{ ,}
\label{Auni}
\end{equation}
where $\Sigma(s)$ is the glueball-glueball self-energy loop function. Note, $A_0\simeq A_0^U$ whenever $A_0$ is small, but sizable loop contributions may occur in general.  
The imaginary part of $\Sigma(s)$ is fixed by the relativistic 2-body phase space of two identical particles:
\begin{equation}
\text{Im}\Sigma(s)=\theta\left(s - 4m_G^2\right)
\frac{1}{2}\frac{1}{16\pi}\sqrt{1-\frac{4m_{G}^{2}}{s}}\label{imloop} \text{ .}%
\end{equation}
The imaginary part of the loop can be used to reconstruct the real part using the dispersion relation by imposing certain constraints. 
The previously discussed requirements of preserving both the pole at $s=m_G^2$ and the logarithmic divergence at $s=3m_G^2$ can be expressed by requiring that
$\Sigma(s=m_{G}^{2})=\Sigma(s=3m_{G}^{2})=0$. Including these subtractions, the loop function takes the form:
\begin{align}
    \Sigma(s)=
    \int_{4m_G^{2}}^{\infty}
\frac{(s-m_{G}^{2})(s-3m_{G}^{2}) \text{ Im} \Sigma(s^{\prime})}{\pi (s^{\prime}-s-i\varepsilon)(s^{\prime}-m_{G}%
^{2})(s^{\prime}-3m_{G}^{2})}ds^{\prime}\text{ .} 
\label{loop}
\end{align}
Interestingly, at threshold the loop takes the value $\Sigma(s=4m_{G}%
^{2})=0.028715$ (independent on $m_{G}$). As a consequence, the $S$-wave unitarized scattering length, called $a_0^U$, takes the form:
 \begin{equation}
a_{0}^{\text{U}}=\frac{1}{32\pi m_{G}}A_{0}^{\text{U}}(s)=\frac{1}{32\pi
m_{G}}\frac{1}{\frac{3\Lambda_{G}^{2}}{92m_{G}^{2}}-0.0028715}.
\end{equation}
One can easily observe that for:
\begin{equation}
\frac{3\Lambda_{G}^{2}}{92m_{G}^{2}}-0.0028715=0,
\end{equation}
the scattering length diverges. Numerically, the critical value for
$\Lambda_{G}$ reads:
\begin{equation}
\Lambda_{G,\text{crit}}=m_{G}\sqrt{\frac{92}{3}\cdot0.0028715}=0.2967m_{G} \text{ .}
\end{equation}
For $m_{G}=1.7$ GeV, the critical value reads $\Lambda_G=0.504$ GeV. In Fig. \ref{figlin} the critical value  $\Lambda_G$ is plotted as function of $m_G$.
The divergence of the scattering
length signalizes the emergence of a glueball-glueball bound state. Such a glueballonium forms whenever the attraction is strong enough, that is when $\Lambda_G < \Lambda_{G,crit}$.
In general, the mass of the glueballonium can be found as a solution of the equation:
\begin{equation}
(A_{0}^{U})^{-1}=A_{0}^{-1}(s)-\Sigma(s)=0.
\end{equation}
We refer to Fig. \ref{figampl}, where the unitarized inverse amplitude is plotted as function of $s$ for three values of $\Lambda_G$. A zero of this quantity (which corresponds to the mass of the glueballonium) is obtained for $\Lambda_G < \Lambda_{G,crit}$. 

\begin{figure}[H]
    \centering
    \includegraphics[width=\linewidth]{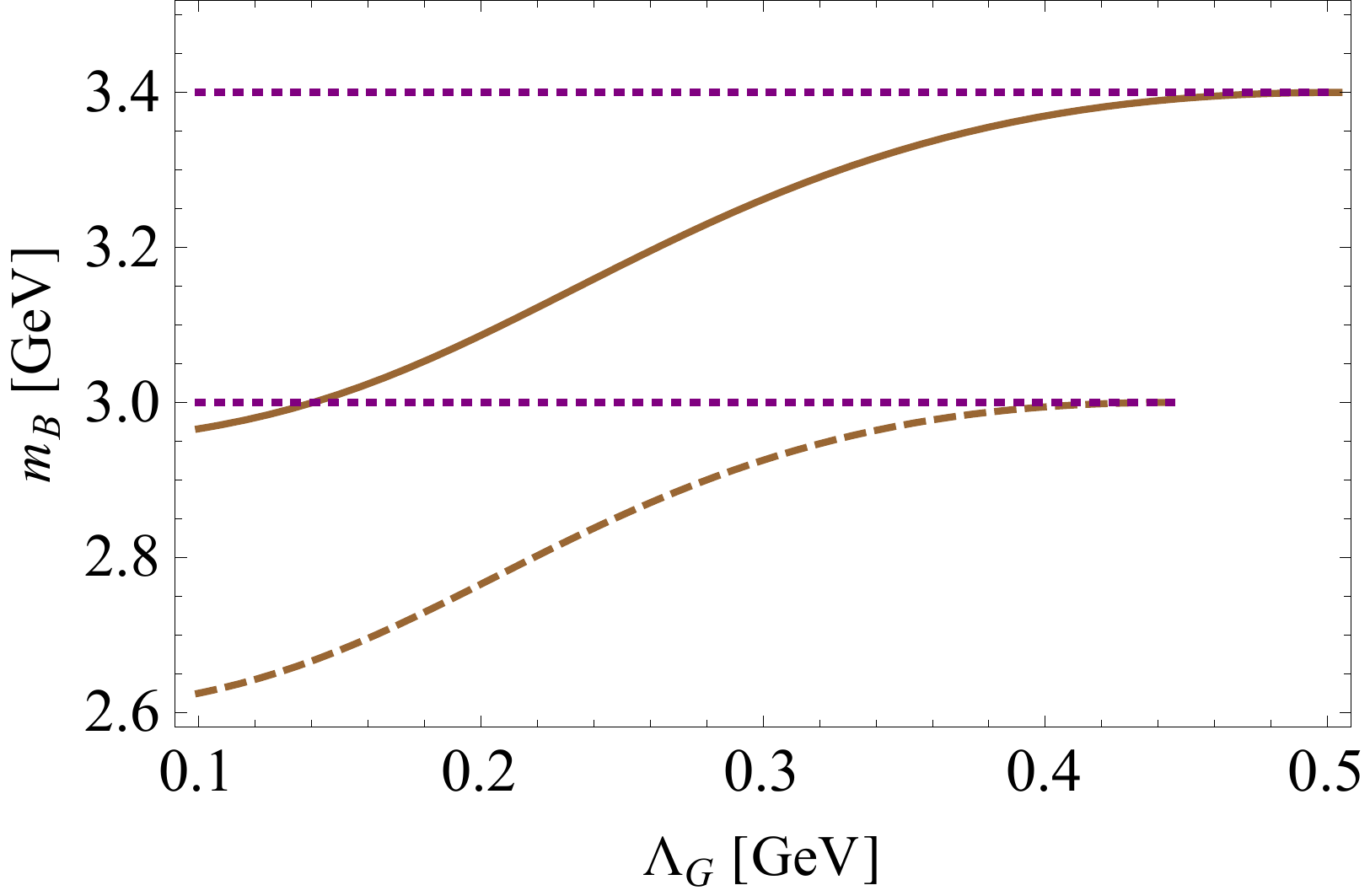}
    \caption{The mass of the glueballonium as function of the parameter $\Lambda_{G}$, using two different glueball masses $m_G=1.5$ GeV (dashed curve) and $m_G=1.7$ GeV (continuous curve).}
    \label{figmass}
\end{figure}

We may summarize the situation as it follows. Depending on the value of the scale: (i) $\Lambda_{G}> \Lambda_{G,crit}$, and (ii) $\Lambda_{G}< \Lambda_{G,crit}$.
For the case (i), the value of the scattering length at threshold is positive and there is no bound state, as the attraction is not strong enough to form it. 
For $\Lambda_{G} \leq \Lambda_{G,crit}$, there is a bound state with mass $m_B \leq 2m_G$. If the scale has a value near the critical one, the pole corresponding to the glueballonium appears near the threshold (and approaches $2m_G$ when $\Lambda_G $ tends to the critical value from below). The mass of the glueballonium ranges between  $\sqrt{3}m_G$ (for $\Lambda_G \rightarrow 0$) and $2m_G$, i.e. between the left-hand cut singularity and the threshold, as it is shown in Fig. \ref{figmass}. In the latter figure, we also present the mass of the glueballonium for two diofferent values of the glueball mass (1.5 and 1.7 GeV, respectively). Note, the glueballonium mass decreases together with $\Lambda_G$ (thus for increasing attraction).
When using the specific value $\Lambda_G \approx 0.4$ GeV obtained in lattice YM, the mass of the glueballonium reads $m_B \approx 3.37$ GeV (for $m_G=1.7$ GeV).

\noindent

\noindent

\section{Conclusions}
In this paper, we used the dilaton potential to study the scattering process between two scalar glueballs. We calculated the scattering length, first at tree level and then in the unitarized version of the theory: this quantity can be evaluated in future Lattice QCD. In this way, the scale $\Lambda_G$, which is important in many effective model of QCD at low energy, could be also fixed by lattice results.

We find that the emergence of a glueball-glueball bound state, that we called glueballonium, is possible. Such a glueballonium is formed when the value of the scale $\Lambda_G$ is below a certain critical value: in the case of $m_G=1.7$ GeV, the critical value of $\Lambda_G$ is around $0.504$ GeV. 
The existence of the glueballonium could be also investigated on the Lattice. An eventual experimental search is also conceivable, if this state turns out to be not too broad. For instance, the planned PANDA experiment \cite{PANDA:2009yku} could search for the existence of this bound state, as it covers an energy range that include the expected mass of the glueballonium (at about 3 GeV). Several ongoing experiments are investigating glueballs \cite{Hamdi:2019dbr,Gutsche:2016wix,Ryabchikov:2019rgx,Marcello:2016gcn,LHCb:2018roe,Belle-II:2018jsg}. As a recent example, the TOTEM and the D\O $~$  collaborations together recently announced the discovery the odderon \cite{TOTEM:2020zzr}, which in turn implies that C-odd-parity glueballs exist. On the other hand, the pomeron implies that C-even parity glueballs (such as the scalar one) exist, see e.g. Refs. \cite{Csorgo:2019ewn,Kirk:1998yi,Albrow:2017gle,Broilo:2020kqg} and refs. therein.

Finally, it would be interesting repeat our work using different unitarization schemes in order to check the dependence of the unitarization on the results. Moreover,  one may study the scattering of other glueballs as well and extend the study at nonzero temperature following the procedure of Ref. \cite{Samanta:2020pez,Samanta:2021vgt}.

\section*{Acknowledgments}
The authors thank A. Pilloni with whom Ref. \cite{Giacosa:2021brl}  was written. 
E.T. acknowledges financial support through the project AKCELERATOR ROZWOJU
Uniwersytetu Jana Kochanowskiego w Kielcach (Development Accelerator of the
Jan Kochanowski University of Kielce), co-financed by the European Union
under the European Social Fund, with no. POWR.03.05.00-00-Z212/18. 
F.G. acknowledges financial support from the Polish National Science
Centre NCN through the OPUS projects no. 2019/33/B/ST2/00613.

\end{multicols}
\medline
\begin{multicols}{2}

\end{multicols}
\end{document}